\documentclass[prb,aps,twocolumn,showpacs,superscriptaddress]{revtex4}
\usepackage{psfig}
\newcommand{\bea}{\begin{eqnarray}}
\newcommand{\beq}{\begin{equation}}
\newcommand{\eea}{\end{eqnarray}}
\newcommand{\eeq}{\end{equation}}
\begin{document}
\title{Roto-vibrational spectrum and Wigner crystallization 
in two-electron parabolic quantum dots}
\author{Antonio Puente}
\affiliation{Departament de F{\'\i}sica, Universitat de les Illes Balears,
E-07122 Palma de Mallorca, Spain}
\author{Lloren\c{c} Serra}
\affiliation{Departament de F{\'\i}sica, Universitat de les Illes Balears,
E-07122 Palma de Mallorca, Spain}
\affiliation{Institut Mediterrani d'Estudis Avan\c{c}ats IMEDEA (CSIC-UIB),
E-07122 Palma de Mallorca, Spain}
\author{Rashid G.~Nazmitdinov}
\affiliation{Departament de F{\'\i}sica, 
Universitat de les Illes Balears, E-07122 Palma de Mallorca, Spain}
\affiliation{Bogoliubov Laboratory of Theoretical Physics,
Joint Institute for Nuclear Research, 141980 Dubna, Russia}

\date{19 August, 2003}

\begin{abstract}
We provide a quantitative determination of the crystallization onset 
for two electrons in a parabolic two-dimensional confinement. 
This system is shown to be well described by a roto-vibrational model,
Wigner crystallization occurring when the rotational motion gets
decoupled from the vibrational one. The Wigner molecule thus formed  
is characterized by its moment of inertia and by the corresponding
sequence of rotational excited states. 
The role of a vertical magnetic field is also considered.
Additional support to the 
analysis is given by the Hartree-Fock phase diagram for the ground state
and by the random-phase approximation for the moment of inertia and 
vibron excitations.  
\end{abstract}


\maketitle

\section{Introduction}

For a low enough electron density, Wigner\cite{W1} predicted that 
electrons should localize creating an ordered spatial structure, 
thenceforth named a Wigner crystal,
that breaks the complete translational symmetry of the homogeneous electron 
gas (also see Ref.\ \onlinecite{Mah}).
Indeed, the formation of the Wigner crystal was observed in two-dimensional
(2D) distributions of electrons on the surface of liquid helium.\cite{2}
A phase transition, induced by the magnetic field, 
from an electron liquid to a crystalline structure has also been reported 
for a 2D electron plasma at a GaAs/AlGaAs heterojunction.\cite{3}

The existence of different phases in quantum dots, 
where a few electrons are confined into a small space, 
has become a topical subject in mesoscopic physics (see, for a recent 
review, Ref.\ \onlinecite{RM}). In fact,
the high controllability of quantum dots suggests that these systems
could provide an attractive
opportunity to achieve the appropriate conditions for localized states. 
It is precisely to stress this controllability that the names 
{\em artificial atoms} and {\em quantum dots} have been coined. 

There is a general persuasion that the Wigner crystallization
in quantum dots, whose localized states are referred to as  
Wigner molecules, should occur at significantly larger densities than in
the 2D bulk. It is based on the argument that in quantum dots 
potential-energy contributions can easily exceed the kinetic
terms and, therefore, electronic motion can be effectively quenched
by manipulating the external confinement and/or an applied magnetic field. 
As for the homogeneous gas, one would expect that in
crystallized states the kinetic energy is solely that of the vibrational 
zero-point motion of the electrons around their mean positions, much smaller
than the interaction (potential) energy. 
Various approaches including ab initio calculations within
diffusion and path integral Monte Carlo methods, Hartree-Fock and spin-density 
functional methods {\em etc} have been applied 
to analyze the onset of the crystallization.\cite{RM}
However, a nonambiguous theoretical result  
that would justify the above conjecture for a zero magnetic field
is lacking. The case with an intense magnetic field is better 
understood since the magnetic field induces an edge reconstruction,
beginning with the appearance of localized vortices on the outer 
region, that ultimately propagates to all the dot for very high 
$B$'s.\cite{RMM,RM} 
 
In the simpler case of a two-electron 2D quantum dot at zero magnetic field, 
Yannouleas and Landman\cite{YL} pointed out that  
the excited-state energies of this system closely follow the 
rotor sequence when the repulsion-to-confinement ratio,
as given by the Wigner parameter $R_W$, is large enough (${\sim}200$). 
This was shown to be a proof of the crystallization of the two electrons
on fixed positions in a reference frame which is rotating.
Quite remarkably, the hypothesized {\em rotating Wigner molecule}
fulfills at the same time the strict symmetry conditions of 
quantum mechanics --circularity in this case-- and the obvious preference 
for opposite positions when repulsion is large enough.
This is a major difference from the above mentioned bulk case
where a Hamiltonian symmetry (translation) is broken by the 
crystallized state. For Wigner molecules, symmetries are preserved 
in the laboratory frame and one must consider an intrinsic (rotating)
frame to 'see' the underlying deformation.        
A similar situation is found 
for particular states of two-electron atoms that have been much 
investigated in physical chemistry (we address the reader to the 
review paper by Berry\cite{Ber}).
For the two-electron quantum dot, however, the crystallization 
condition from Ref.\ \onlinecite{YL}, $R_W\sim 200$, looks disappointing since 
it seems unrealistic to achieve such a value experimentally.

Although the exact ground-state wave function of the two-electron artificial 
atom can be obtained, at least numerically, it may seem 
paradoxical that one also needs the excited states in order to ascertain 
the existence of a crystallization. In fact, this inability to 
disentangle the system's intrinsic structure from its full wave function 
in a clear way can be taken as a weakness of the ab initio, symmetry 
preserving, approaches. In general, even in cases when the exact 
ground- and excited-state wave functions and energies are known, an intrinsic 
deformation can only be inferred by comparing with the result of 
simpler models in which either symmetries are relaxed or the intrinsic 
structure is imposed. A clear example of the former approach is given by the 
unrestricted Hartree-Fock (HF) method for the ground state\cite{Koo,Yan} 
followed by the random-phase approximation (RPA) for excitations.\cite{rpa} 
Conversely, the roto-vibrational model of Wendler {\em et al.}\cite{Wen}
for two electrons in a ring could be included in the latter category.

One should be aware that when symmetries are relaxed, as in the Hartree-Fock
approach, artifacts or unphysical properties may appear. 
In a recent contribution Reusch and Grabert\cite{RG} 
discussed the validity of the latter,
drawing special attention to the caution with which one must take 
Hartree-Fock predictions on symmetry breaking, in agreement with the 
results presented below. Therefore, a complete physical understanding requires 
both exact results and model solutions. This way the system's intrinsic deformations 
are physically understood while, at the same time, artifacts can be safely
discarded. 
A paradigmatic case where the proposed analysis can be performed is given by 
the two-electron 2D parabolic quantum dot. The separation of center-of-mass and 
relative coordinates along with the circular symmetry restriction allows the 
reduction of the multidimensional 
Schr\"odinger equation to just a radial one, easily solvable numerically. 
On the other hand, the Hartree-Fock and RPA solutions without any symmetry 
restriction can also be obtained. A most convenient basis for this latter 
calculations is given by the Fock-Darwin orbitals in terms of which one can 
analytically develop much of the required algebra. 
It is our aim in this work to determine the crystallization
onset of two-electron parabolic dots by recourse to the three different
approaches referred to above; namely, (a) an analytical roto-vibrational model,  
(b) a numerical solution of the Schr\"odinger equation, and 
(c) symmetry unrestricted Hartree-Fock and random-phase approximations.

Hereafter, we refer to the solution of the Schr\"odinger equation
for the two-electron parabolic dot as the {\em exact} solution. It should 
be pointed out that, as shown by Taut,\cite{Taut} this Schr\"odinger equation 
is analytically solvable only for particular confinement/interaction 
strengths. For general values of this quantity a numerical treatment
is required. As mentioned above, the most straightforward one is an 
integration of the radial equation\cite{YL,Zhu} but, nevertheless, other 
methods such as diagonalization in a basis\cite{Pfan,Wag} and 
the Monte Carlo method \cite{Ped,Loz} have also been applied.  
One of us has used the so-called oscillator representation method,
perturbatively treating the residual interaction, to derive analytical 
expressions for the energy levels.\cite{Din}

The paper is organized as follows. Section II introduces 
the magneto-parabolic units that allow one to trace the evolution of ground and 
excited states of artificial atoms at various conditions. 
An analytical roto-vibrational model for the two-electron parabolic quantum 
dot is described in Sec.\ III. Section IV provides details of 
our numerical calculation of exact solutions and compares these solutions 
with those of the roto-vibrational model.
Section V analyzes the reliability of 
the Hartree-Fock and RPA results for the present system. 
A short summary is finally drawn in Sec.\ VI.

\section{Magneto-parabolic unit system}

We consider two electrons with a Coulomb 
interaction between them. The electrons move in the $xy$ plane 
where a circular parabolic confinement induces the formation of an electron island. 
The system is also subject to an external magnetic field applied in the
vertical direction ($z$). The full Hamiltonian thus
reads
\begin{eqnarray}
\label{eqH}
{\cal H} &=& \sum_{i=1,2}{\left[
{1\over 2 m} \left({\bf p}+\frac{e}{c}{\bf A}\right)^2 + 
\frac{1}{2} m \omega_0^2 r^2
\right]_i}+ {e^2\over \kappa\, r_{12}}\nonumber\\
&+& g^* \mu_B B S_z\; .
\end{eqnarray}
In Eq.\ (\ref{eqH}), $m$, $\kappa$, and $g^*$ are the electron's
effective mass, the dielectric constant, and the effective gyromagnetic 
factor, respectively, and we have used planar polar coordinates 
($r^2=x^2+y^2$). The two contributions within the square brackets are,
respectively, the generalized kinetic energy in terms of the vector 
potential ${\bf A}$, and the external confinement.
Within the so-called symmetric gauge one has 
${\bf A}(x,y)=B/2(-y,x)$. The next contribution is the 
Coulomb repulsion and, finally, the last term is the Zeeman energy 
involving the total spin 
operator $S_z$ and the universal Bohr's magneton $\mu_B=e\hbar/2m_ec$.

It is well known that in the chosen gauge the magnetic field contributions 
can be recast into the form of an effective parabolic confinement of
frequency $\Omega=\sqrt{\omega_0^2+\omega_c^2/4}$, where 
$\omega_c=eB/mc$ is the cyclotron frequency, and an additional angular-momentum-dependent 
term $(\hbar\omega_c/2)\ell_z$ (cf.\ Refs.\ \onlinecite{Hawrylak,Chak}). 
The magneto-parabolic units (mpu) we shall use consist of
taking $\hbar\Omega$ as the energy unit and $\ell_\Omega\equiv\sqrt{\hbar/(m\Omega)}$
as the length unit. In addition, one also imposes $\hbar$ as angular momentum 
unit which, in turn, fixes the time unit $\tau_\Omega=1/\Omega$. Summarizing in the 
standard abuse of notation we may write $\hbar=\Omega=\ell_\Omega=1\; {\rm mpu}$. 
This is a natural choice for magneto-parabolic confinements and it allows
one to express the spatial part of the Hamiltonian in terms of only two 
adimensional parameters, namely,
\begin{eqnarray}
\label{eq2}
R_{\it mp} &=& {e^2/(\kappa\ell_\Omega)\over \hbar\Omega}\; ,\\
\label{eq3}
W_{\it mp} &=& {\omega_c\over \Omega} \;.
\end{eqnarray}
Note that $R_{\it mp}$ and $W_{\it mp}$ give, respectively, the ratios of Coulomb 
interaction strength and cyclotron frequency to effective confinement.
 In the absence of a magnetic 
field $R_{\it mp}$ coincides with the so-called Wigner parameter
$R_W$ of Ref.\ \onlinecite{YL}. 
Also note that $W_{\it mp}$ has a maximal value $W_{\it mp}=2$  that
corresponds to a zero confinement $\omega_0=0$.
We also mention that Reusch and Grabert 
used these parameters in their recent Hartree-Fock calculations.\cite{RG}

\begin{figure}[t]
\centerline{\psfig{figure=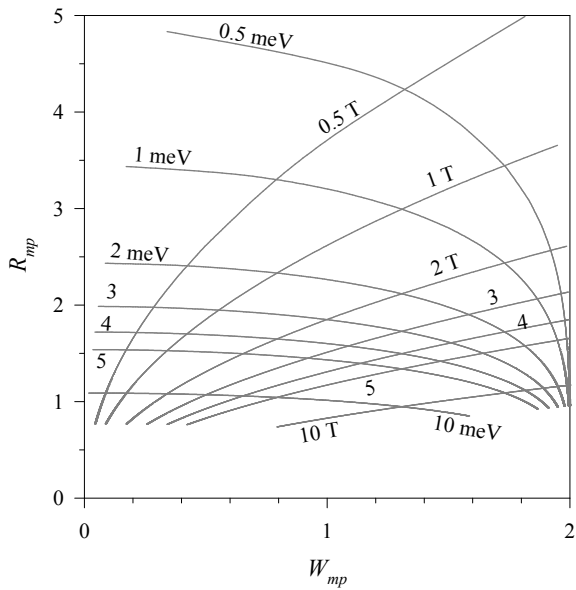,width=3.in,clip=}}
\caption{Equivalence between the mpu pair of adimensional 
paramaters $(R_{\it mp},W_{\it mp})$ and the physical values
of the external parabolic strength and magnetic field $(\hbar\omega_0,B)$. 
The bulk GaAs effective mas $m=0.067 m_e$ and dielectric constant
$\kappa=12.4$ have been assumed.}
\label{fig1}
\end{figure}

Within the mpu system the Hamiltonian reads
\begin{eqnarray}
\label{eqmpH}
{\cal H} &=& \sum_{i=1,2}{\left[
-\frac{1}{2}\nabla^2 + \frac{1}{2} r^2+ \frac{W_{\it mp}}{2} \ell_z
\right]_i}+ \frac{R_{\it mp}}{r_{12}}\nonumber\\
&+& \frac{g^* m^*}{2} W_{\it mp} S_z\; ,
\end{eqnarray}
where $m^*$ is the adimensional ratio of effective to bare electron
mass, i.e., $m^*=m/m_e$.
The passage from the mpu system, with a given $R_{\it mp}$ and 
$W_{\it mp}$, to physical units requires the knowledge of the 
effective mass $m$ and the dielectric constant $\kappa$. More specifically,
with fixed $m$ and $\kappa$
one can invert Eqs.\ (\ref{eq2}) and (\ref{eq3}) for the effective confinement
$\Omega$ and cyclotron $\omega_c$ frequencies or, equivalently, for the external 
confinement $\omega_0$ and the magnetic field $B$. In terms of these 
results
the physical values of the energy ($\hbar\Omega$) and length ($\ell_\Omega$) 
units are readily found. 

Figure \ref{fig1} shows the equivalence between
the adimensional parameters $(R_{\it mp},W_{\it mp})$ of the mpu Hamiltonian
and the physical values of $(\hbar\omega_0,B)$ for the case of a GaAs host 
semiconductor having $m=0.067m_e$ and $\kappa=12.4$.  
The advantage of working with the mpu system becomes obvious when realizing that, 
with unit redefinition, the same numerical results can be equally applied to 
a variety of confinements, magnetic fields, and material parameters (effective 
mass and dielectric constants). Therefore,
the model acquires a certain degree of universality. We also expect that 
quantum dot properties such as scaling laws or phase diagrams will be better 
displayed in terms of the adimensional mpu parameters.

\section{A roto-vibrational model}

Taking Eq.\ \ref{eqmpH} as a starting point and introducing the standard 
center of mass $(R,\Theta)$ and relative $(r,\theta)$ coordinates
it is well known that the Hamiltonian separates and, therefore, that the 
wave function factorizes. The center of mass (CM) problem is that of a 
single particle in a harmonic potential and magnetic field, having an
analytical solution in terms of Fock-Darwin orbitals and energies
$\varepsilon^{\rm (CM)}_{NM}=2N+|M|+1+M W_{\it mp}/2$, 
with $(N,M)$ the 
radial and angular momentum center-of-mass quantum numbers. Focusing next on 
the relative problem, one introduces the wave function 
$e^{im\theta} u_{nm}(r)/\sqrt{r}$ having good $\ell_z$ angular momentum ($m$), 
and an additional quantum number $n$ whose meaning will be clarified below. 
The equation for the unknown $u_{nm}(r)$ reads
\begin{equation}
\label{eqrel}
u_{nm}''+\left[
\tilde\varepsilon^{\rm (rm)}_{nm}-\left(\frac{1}{4}r^2+\frac{R_{\it mp}}{r}
+ {m^2-1/4\over r^2}\right)
\right] u_{nm} = 0 \; ,
\end{equation}
where we have defined 
$\tilde\varepsilon^{\rm (rm)}_{nm}=\varepsilon_{nm}^{\rm (rm)}-m W_{\it mp}/2$
in terms of the relative-motion energy $\varepsilon_{nm}^{\rm (rm)}$
and the mpu parameter $W_{\it mp}$.

Equation \ (\ref{eqrel}) will be the basis of our roto-vibrational
model. Note that it resembles a Schr\"odinger one-dimensional equation
with an effective potential 
\begin{equation}
\label{eq7}
V_{\it eff}(r)=\frac{1}{4}r^2+\frac{R_{\it mp}}{r}
+ {m^2-1/4\over r^2}\; 
\end{equation}
that includes the rotational motion term $\sim m^2/r^2$ characterized
by the angular momentum quantum number $m$.
We can expect a rigid-rotor behaviour if $V_{\it eff}(r)$
has a deep minimum at a particular value $r=r_0$.
When this occurs the situation resembles that of diatomic
molecules like H$_2$, where the potential well 
for nuclear motion is described by the Morse potential 
(see Ref.\ \onlinecite{BJ}). 
In the present case the effective potential indeed has a minimum although
it is in general rather shallow. This property is responsible for
the coupling between rotation and vibration or, equivalently, for the  
{\it floppiness} of the rotating molecule mentioned in Ref.\ \onlinecite{YL}.

The minimum condition on $V_{\it eff}(r)$ yields the rotor radius 
from
\begin{equation}
\label{eq8}
\frac{r_0}{2}-\frac{R_{\it mp}}{r_0^2}-{2(m^2-1/4)\over r_0^3} =0\; .
\end{equation}
Neglecting the third contribution on the left-hand-side, an assumption
that will be valid for large enough $r_0$,
one finds the asymptotic law $r_0\approx (2R_{\it mp})^{1/3}$.
Now, expanding to second order around $r_0$ we approximate
\begin{eqnarray}
V_{\it eff}(r) &\approx& V_{\it eff}(r_0) + 
\frac{1}{2}\left(\frac{3}{2}+2{m^2-1/4\over r_0^4}\right)
(r-r_0)^2\nonumber\\
&=& {\it const.} + \frac{1}{2}k(r-r_0)^2\; ,
\end{eqnarray} 
a result that, when substituted into Eq.\ (\ref{eqrel}) for the round 
parentheses, leads immediately to the analytical prediction\cite{ana}
\begin{eqnarray}
\tilde\varepsilon^{\rm (rm)}_{nm} &=& \frac{1}{4}r_0^2+\frac{R_{\it mp}}{r_0}
+ {m^2-1/4\over r_0^2} \nonumber\\
\label{eq9}
&+& \left(n+\frac{1}{2}\right)
\sqrt{3+4{m^2-1/4\over r_0^4}}\; .
\end{eqnarray}
 
The result embodied by Eq.\ (\ref{eq9}) has a clear physical interpretation.
It contains a rotor-like contribution, $\sim m^2/(2{\cal J})$, with a moment 
of inertia given by ${\cal J}=r_0^2/2$ mpu, and a vibrational one characterized 
by a quantum number $n$. The vibrational frequency 
$\omega_{\it vib}=\sqrt{k/\mu}$ ($\mu=1/2$) is given by the last square-root 
factor. Similar to atomic molecules, there is roto-vibrational 
coupling, since the vibration frequency depends on $m$ and, in addition, centrifugal 
distortion since $r_0$ also depends on $m$. For large enough values of $R_{\it mp}$,
implying large $r_0$ and therefore small average densities,
the centrifugal distortion disappears and one has $r_0\approx(2R_{\it mp})^{1/3}$
for all $m$'s.
In this limit rotational terms become negligible, as well as roto-vibrational
ones. Thus, Eq.\ (\ref{eq9}) reduces to a simple $m$-independent asymptotic 
expression
\begin{equation}
\label{asympt}
\tilde\varepsilon^{\rm (rm)}_{n} = \frac{3}{2^{4/3}} R_{\it mp}^{2/3}
+\sqrt{3}\left(n+\frac{1}{2}\right)\; .
\end{equation}

When adding the magnetic field, the roto-vibrational energy becomes
\begin{eqnarray}
\varepsilon^{\rm (rm)}_{nm}&=& \tilde\varepsilon^{\rm (rm)}_{nm}+
m W_{\it mp}/2\nonumber\\
&\sim& {(m+W_{\it mp} r_0^2/4)^2\over r_0^2}+ \left(n+\frac{1}{2}\right) \omega_{\it vib}\; ,
\end{eqnarray}
in agreement with the expectations from Ref.\ \onlinecite{Wen} for two interacting 
electrons in a quantum ring. \cite{Pet} 
It is worth stressing that since the $W_{\it mp}$ dependence only amounts
to an energy shift of the Eq.\ (\ref{eqrel}) eigenvalue, the radial function 
$u_{nm}(r)$ does not depend on $W_{\it mp}$. Therefore, one may conclude that 
the roto-vibrational properties are not affected by magnetic fields, for a
fixed $(n,m)$ state. Of course, since the energy shift varies for different 
states, the magnetic field will modify the ordering of energy levels. For instance,
the level crossings as a function of $W_{\it mp}$ will cause the ground state
to have a nonvanishing $m$-value. This actually explains the buildup 
of increasing permanent currents in the dot's ground state. 
 
The results from this section will be validated by comparing with the exact 
ones below. The roto-vibrational model presented here allows one to determine
the crystallization onset from the criterion that rotation and vibration
motions decouple when intrinsic-frame electron localization sets in. 
Conversely, when the coupling is strong
the system could be represented by either a vibrating rotor or a 
rotating vibrator and, therefore, the situation can not be clearly resolved.
It is also worth stressing that the roto-vibrational model describes
all possible excitations of the relative-motion problem. For this 
particular system, this amounts to a description of all the excitations
since the center-of-mass and spin degrees of freedom can be 
analytically integrated out.
  
\begin{figure}[t]
\centerline{\psfig{figure=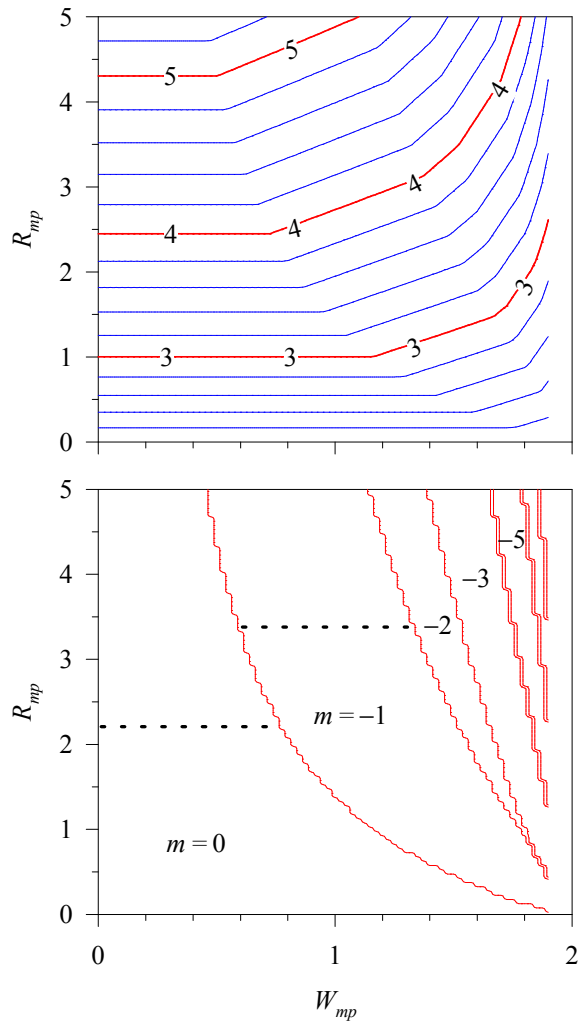,width=3.in,clip=}}
\caption{(Color online) Upper panel: Ground state energy in mpu's for the two electron
dot. Lower panel: Ground state relative angular momentum. Note that 
since $M=0$, relative and total angular momentum coincide.
Even-$m$ (odd-$m$) regions correspond to singlet (triplet) ground states.
Results for $1.9\le W_{\it mp}\le 2$ are not shown due to the 
excessively large variations of the computed quantities in this region.
The dotted lines separate in each domain with a given $m$ the 
crystallized (above) from the non crystallized (below) phases using
the criterion of roto-vibrational coupling below 3 {\%} (See subsect. IV.B).}
\label{fig2}
\end{figure}

\begin{figure}[t]
\centerline{\psfig{figure=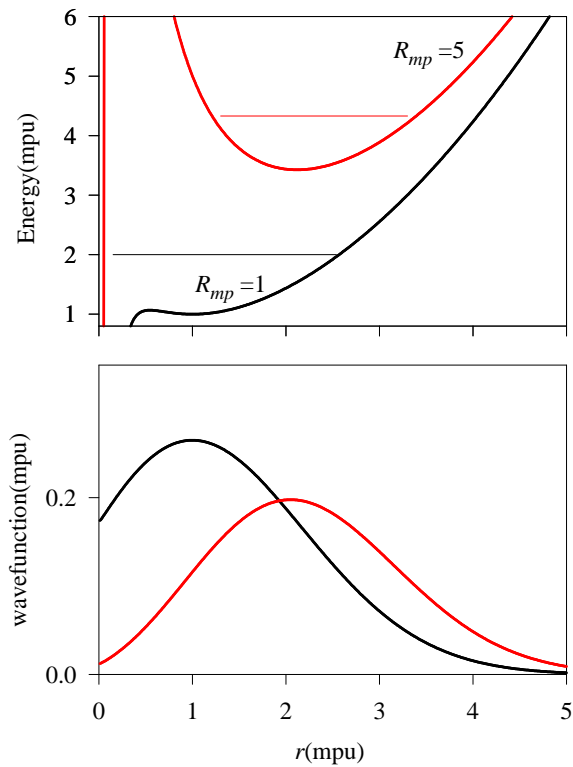,width=3.in,clip=}}
\caption{(Color online) Upper panel: Effective potential defined by Eq.\ (\ref{eq7})
for $m=0$ and two different values of $R_{\it mp}$. The relative-motion 
eigenenergy $\tilde\varepsilon_{00}$ is given in each case by the 
horizontal line. Lower panel: radial eigenfunction $u_{00}(r)/\sqrt{r}$ for the same  
two potentials of the upper panel.}
\label{fig3}
\end{figure}

\section{Onset of Wigner crystallization}

\subsection{Exact solutions}

We have solved the radial equation for the relative problem, 
Eq.\ (\ref{eqrel}), numerically. Several standard approaches to eigenvalue
problems with boundary conditions can be used for 
this purpose. Specifically, we have applied the so-called matching 
method where one integrates with the Numerov or Runge-Kutta algorithms from 
the origin $r=0$ outwards assuming an
$r^{|m|+1/2}$ behaviour. Additionally, imposing an exponential decay law
$\exp{(-r^2/4)}$ for large 
$r$, inwards integration is performed and the two solutions are required to
match at an intermediate $r$-point. To ascertain the numerical result for the 
eigenvalue, as a control we also used the method of 'node-counting', where
only outwards integration is performed and the eigenvalue is found from
the condition that $u_{nm}(r)$ increases by one the number of radial nodes
from the required value when the energy exceeds  the correct eigenvalue by 
an infinitesimal amount. In principle, the node-counting method assures the correct
boundary condition at $r\to\infty$ automatically. However, in practice, there always
remains a small difference between the numerical and exact eigenvalues,
responsible for a deviation from the exponential decay from some
(large) $r$ onwards. In spite of this possible difference in asymptotic
behavior, the two methods (matching and nodecounting) 
provide to a high accuracy the same eigenvalue.

With the above numerical methods an exploration of a part of the 
$R_{\it mp}-W_{\it mp}$ plane
has been performed. Figure \ref{fig2} summarizes the results for ground-state energy and
angular momentum. Note that the ground state always has $(N,M)=(0,0)$ for 
the CM quantum numbers and that, because of symmetry, even (odd) $m$ states
are associated with singlet (triplet) total spin. Clear
singlet-triplet oscillations, as studied in the literature 
(cf.\ Ref.\ \onlinecite{Wag}), are seen in the lower 
panel of Fig.\ \ref{fig2}. It is also worth mentioning that the ground-state energy 
contour lines are piece-wise linear as a result of the simple dependence
on the $W_{\it mp}$ parameter. Indeed, the energy is totally $W_{\it mp}$-independent 
for $m=0$ and a fixed $R_{\it mp}$.

Figure \ref{fig3} displays the radial wave functions $u_{00}(r)/\sqrt{r}$ 
for two different values
of $R_{\it mp}$ as well as the corresponding effective potentials $V_{\it eff}$ 
defined by Eq.\ (\ref{eq7}). We note that, in agreement with the discussion of 
the preceding section, when increasing $R_{\it mp}$
the effective-potential minimum moves outwards and it
effectively binds the lower states to its neighborhood. When this 
occurs the radial probability is strongly quenched
at small $r$ (lower panel) and the scenario indeed resembles the familiar one from 
the physics of diatomic molecules. A more detailed comparison of the 
roto-vibrational model with the exact results will be done in the 
next subsection.
  
\begin{figure}[t]
\centerline{\psfig{figure=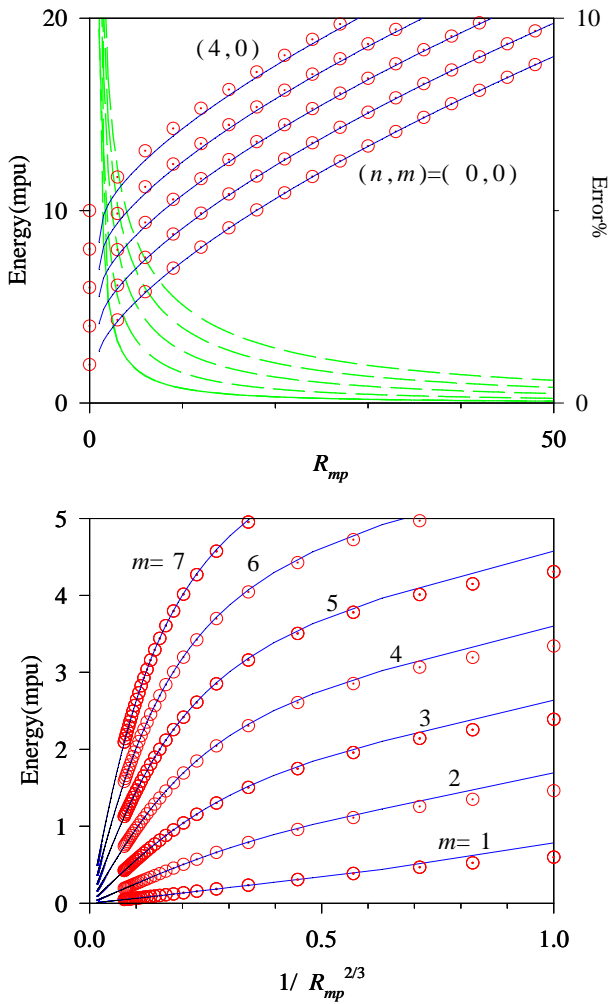,width=3.2in,clip=}}
\caption{(Color online) Upper panel: $\tilde\varepsilon_{nm}$ exact energies 
and the analytical prediction of the 
roto-vibrational model (thin solid curves). The dashed lines 
give the relative error in percentage (right scale) with bottom
to top curves corresponding to $n=0$, 1, \dots  4, respectively.
Lower panel: Excitation energies of the $(0,m)$ states, i.e, 
$\tilde\varepsilon_{0m}-\tilde\varepsilon_{00}$ from the exact (data)
and analytical model (curves).}
\label{fig4}
\end{figure}

\subsection{Crystallization criterion} 

The validity of the roto-vibrational model presented in Sec.\ III is 
proved by the results of Fig.\ \ref{fig4}. As shown in the upper
panel, the error of the analytical prediction for $\tilde\varepsilon_{nm}$
is important only when $R_{mp}$ is small. As a matter of fact, for $R_{mp}>2$ 
the discrepancy for $\tilde\varepsilon_{00}$ is always below 2{\%}, 
even with the asymptotic expression Eq.\ (\ref{asympt}).
Note that the relative errors slightly increase with increasing $n$ (dashed curves),
although the analytical approximation can still be considered quite good.   
The analogous comparison for $m>0$ (not shown) yields smaller
relative errors than those for $m=0$.
The lower panel of Fig.\ \ref{fig4} analyzes the excitation energies
as measured from $\tilde\varepsilon_{00}$, again showing 
an excellent agreement between the exact results and the ones obtained 
with the analytical model, with small deviations only at small $R_{\it mp}$. 
Based on the asymptotic expression, Eq.\ (\ref{asympt}), we have chosen 
$1/R_{\it mp}^{2/3}$ as an independent variable in order 
to better display the linear behavior associated with the rigid rotor 
at large $R_{mp}$.

\begin{figure}[t]
\centerline{\psfig{figure=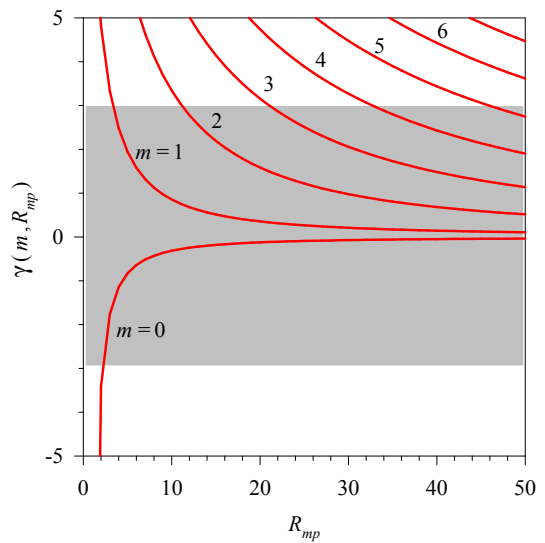,width=2.75in,clip=}}
\caption{(Color online) Roto-vibrational coupling as defined in Eq.\ (\ref{eqrvc}).
The shaded region indicates the crystallized phase with the criterion
$\gamma\le 3$.}
\label{fig5}
\end{figure}

We shall rely on the high accuracy of the roto-vibrational model to
provide a quantitative measure of the crystallization onset. Our 
criterion will be the following: the two-electron parabolic system
is assumed to be crystallized as a rotating Wigner molecule when the 
roto-vibrational coupling falls below a given percentage (typically chosen 
as 2 or 3 {\%}). 
The roto-vibrational coupling is defined as
\begin{equation}
\label{eqrvc}
\gamma(m) = 100 {\omega_{\it vib}(m;R_{\it mp})-\sqrt{3}\over \sqrt{3}}\; ,  
\end{equation}
where $\omega_{\it vib}$ is the square-root factor in Eq.\ (\ref{eq9})
and $\sqrt{3}$ is the limit of this quantity for $R_{\it mp}\to\infty$.
Using a {3\%} 
condition the crystallization onset for each angular momentum 
is given by the $R_{\it mp}$ value where each $m$ curve of Fig.\ \ref{fig5}
enters the shaded region. 
Note that the crystallization onset
moves towards higher values as the angular momentum is increased, 
reflecting the property that roto-vibrational coupling is stronger
for high-$m$ states. It is also worth mentioning that, since not all 
$m$-states are simultaneously crystallized, in practice the {\em rotational bands} 
will gradually degrade for increasing $m$'s as a consequence of
the roto-vibrational coupling.
In agreement with the bulk gas situation, the crystallized states with 
the proposed criterion are characterized by having a potential
energy that largely exceeds the kinetic one, as can be easily checked from
Eq.\ (\ref{eq9}).

The crystallization properties of $m$ and $-m$ states are identical
since one can easily check that 
$u_{nm}(r;R_{\it mp})=u_{n-m}(r;R_{\it mp})$. Therefore, taking
into account the variations in ground state angular momentum we
can draw the boundaries for Wigner crystallization in the 
$R_{\it mp}-W_{\it mp}$ plane, i.e., the crystallization
phase diagram (see the lower panel of Fig.\ \ref{fig2}). Of course, if instead
of a 3{\%} threshold for roto-vibrational decoupling 
one chooses a different value the crystallization onset will vary, 
although as shown in Fig.\ \ref{fig5}, for low $m$'s the crystallization
is not crucially dependent on the precise percentage in the range
2 to 4{\%}. 
Actually, it should be more appropriate to speak of crystallization
onset for a given percentage of roto-vibrational decoupling than of an 
absolute value.

\begin{figure}[t]
\centerline{\psfig{figure=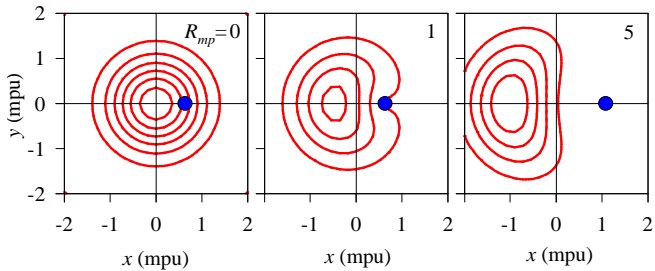,width=3.4in,clip=}}
\caption{(Color online) Contour plots of the ground state two-electron 
wave function $|\Psi({\bf r}_1,{\bf r}_2)|^2$ for the fixed value of ${\bf r}_2$ 
shown by a solid symbol. A value of $W_{mp}=0$ and the displayed
$R_{mp}$ have been used for the different panels. For $R_{mp}=0$,
${\bf r}_2$ has been arbitrarily fixed on the $x$ axis, while for $R_{mp}>0$
it has been placed at the distance inferred from the asympotic law
$x_2=r_0/2$ (see text).}
\label{fig6}
\end{figure}

It should be noted that the onset of Wigner crystallization 
may be studied by means of the conditional probability distribution (CPD,
cf.\ Refs.\ \onlinecite{Yan,Mik}) for finding one electron
at positon ${\bf r}_1$ given that the second electron is at position 
${\bf r}_2$. As soon as the interaction is switched on,
the CPD, which for a system with only two electrons is just 
the modulus squared of the wave-function 
$|\Psi({\bf r}_1,{\bf r}_2)|^2$, exhibits the 
formation of a molecular-like state (see Fig.\ \ref{fig6}). 
It is difficult, however, to use this measure alone as a 
conclusive evidence of the formation of the Wigner molecule
in two-electron dots.
The difficulty arises from the fact that
even weak interactions, for which we should not expect a 
crystallization, yield the formation of a hole around
the electron at ${\bf r}_2$ (the correlation hole) 
and a maximum at ${\bf r}_1=-{\bf r}_2$.
Indeed, as seen from Fig.\ \ref{fig6}, the results for
$R_{mp}=1$ hint at a molecular state even though
according to our analysis the roto-vibrational coupling is still strong 
and this state is not associated
with a crystallized phase (see Fig.\ \ref{fig5}). 
When $R_{mp}$ increases the depletion of the 
CPD around the fixed electron becomes much stronger, which is in qualitative 
agreement with the crystallization trend from our analysis. 

\section{Hartree-Fock and RPA approaches}

In this section we discuss the results obtained within 
the symmetry unrestricted Hartree-Fock method for the ground state 
and the corresponding RPA for excitations. 
We recall that the HF and RPA approaches were devised
for the analysis of many-body systems. 
Therefore, their application to a two-electron
quantum dot is merely an exploratory approach to the
qualitative features of the Wigner crystallization rather than a quantitative 
description of the above exact results. 

We have solved the HF problem in 
the Fock-Darwin basis which diagonalizes the square bracket in 
Eq.\ (\ref{eqH}), namely 
$\{\, |a \eta\rangle;\, a=1,\dots {\cal N};\, \eta=\uparrow,\downarrow\, \}$, 
where $a$ labels the orbital part
and $\eta$ the spin. 
Our basis has been optimized such a way that we consider the 70 lowest 
Fock-Darwin states, of the non-interacting energy level scheme,  
for a chosen value of the magnetic field.
An arbitrary single-particle orbital $|i\rangle$ is then expanded as $|i\rangle = 
\sum_{a\eta} B^{(i)}_{a\eta}\, |a\eta\rangle$. In the chosen basis, 
the HF equations are written as a system of nonlinear eigenvalue equations 
for the matrix of $B$ coefficients (see, for instance, details in Ref.\ 
\onlinecite{rpa}).

We have imposed good $s_z$ HF orbitals leaving totally unspecified the 
remaining spatial symmetries. Note that the Slater determinant built with these 
single-particle orbitals will be an eigenstate of the total $S_z$ operator 
but not, in general, of ${\bf S}^2$. This, as we shall see in the results,
is intimately connected with the prediction of broken circular
symmetry.

\begin{figure}[t]
\centerline{\psfig{figure=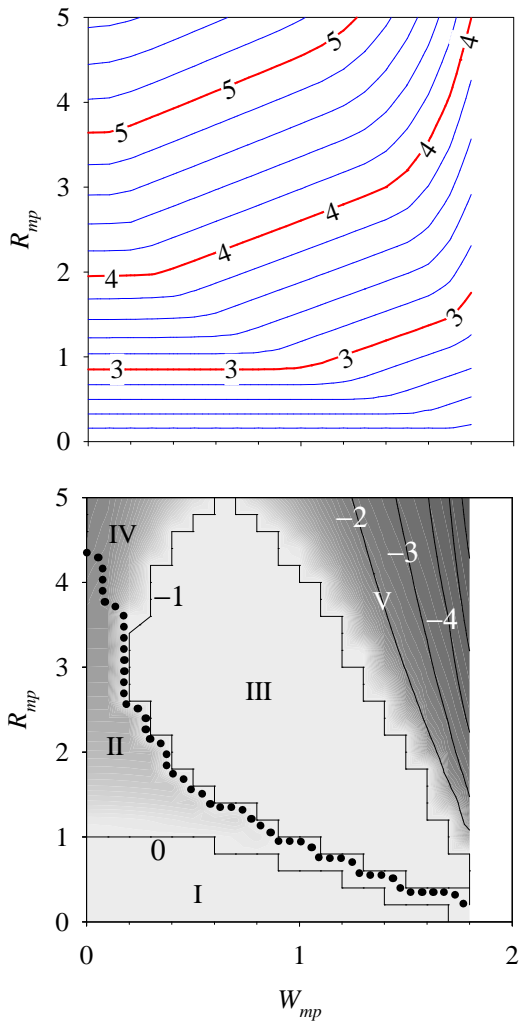,width=2.75in,clip=}}
\caption{(Color online) Upper panel: HF ground state energy in mpu's for the two electron
dot. Lower panel: Regions ({\small I} to {\small V}) of the HF ground state
phase diagram.
The dotted curve separates total $S_z=0$ (below) and $S_z=1$ (above) configurations. 
The gray scale denotes the space symmetry of the HF solution,\cite{not1} 
from circular (light gray) to strongly deformed (dark gray) configurations.
Labeled contour lines display $\langle L_z\rangle$. Note that in the 
broken-symmetry regions ({\small II}, {\small IV} and {\small V})
this latter quantity varies continuously between the integer boundaries.
Results for $W_{\it mp}>1.8$ are not shown due to the difficulty in 
determining the single-particle basis when the Fock-Darwin states become 
quasi-degenerate.}
\label{fig7}
\end{figure}

In Fig.\ \ref{fig7} we show the HF phase diagram in the $R_{\it mp}-W_{\it mp}$ plane
(lower panel) and the corresponding total energies (upper panel). The total 
energies resemble those obtained in the exact treatment, with approximate 
piecewise linear regions between orbital angular momentum transition lines. 
As expected the actual values at a given point in the diagram lie slightly 
above the corresponding exact results. 
In the lower panel different gray regions reflect a measure\cite{not1} of the 
deviation from circularity of the ground state density, with the lightest intensity 
corresponding to a circular (nonbroken symmetry) solution and more intense gray 
levels to noncircular (broken symmetry) results. 
The contour lines show the total orbital angular momentum $L_z$.
Regions {\small I} and {\small III} are of circular symmetry and for them 
$L_z$ has a good quantum number, taking the values 0 and -1, respectively. For the rest 
of the diagram (regions {\small II}, {\small IV} and {\small V}) the contour 
lines only indicate the expectation value of $L_z$ but, since 
the mean field in not circularly symmetric, this is no longer a 
good quantum number.  The dotted curve separates the states having 
total spin projection $S_z=0$ (below) from $S_z=1$ (above).

For $S_z=0$ configurations, the HF method predicts a broken-symmetry solution when  
$R_{\it mp}$ exceeds a value $\simeq 1$, somehow below the onset of crystallization obtained 
in Sec.\ IV ($R_{\it mp} \simeq 2.2$). In this region ({\small II}) the HF solution is indeed
a mixture of singlet and triplet states, as can be verified by computing the 
total spin dispersion $\Delta{\bf S}^2$. The corresponding spatial density is built 
from two opposite and localized single electron orbitals.
It is instructive to compare the HF solutions in region {\small II} with those obtained 
using a total-spin conserving ansatz
\begin{eqnarray}
\label{sng}
\Psi_{\it sng}({\bf r}_1,{\bf r}_2,\eta_1,\eta_2) &=&
\phi({\bf r}_1)\phi({\bf r}_2) \ \chi_{\it sng}(\eta_1,\eta_2)\\
\Psi_{\it trp}({\bf r}_1,{\bf r}_2,\eta_1,\eta_2) &=&
{\cal A}[\phi_1({\bf r}_1),\phi_2({\bf r}_2)] \nonumber\\
&& \times \chi_{\it trp}(\eta_1,\eta_2) 
\label{trp}
\end{eqnarray}
for singlet ($\Psi_{\it sng}$) and triplet ($\Psi_{\it trp}$) states, where 
${\cal A}[\phi_1({\bf r}_1),\phi_2({\bf r}_2)]$ denotes the antisymmetrized product 
of the two orbitals $\phi_1$ and $\phi_2$ while the $\chi$'s are the well 
known singlet and triplet spin states.

As a sample result the total energies obtained for 
$(W_{\it mp},R_{\it mp})=(0,2)$ are $E_{\it exact}=3.720$, $E_{HF}=4.034$, 
$E_{\it sng}=4.185$ and $E_{\it trp}=4.168$; i.e., by requiring total ${\bf S}^2$ 
conservation the mean field energy raises considerably. In addition, while the HF 
solution breaks circular symmetry, both $\Psi_{\it sng}$ and $\Psi_{\it trp}$
yield circular densities because of the spatial dependence of
the $\phi$'s in Eqs.\ (\ref{sng}) and (\ref{trp}). 
We have also checked that these results are equivalent to those 
obtained using the Lipkin-Nogami projection method \cite{Lip,Nog} for the
effective Hamiltonian $H_{\rm eff}={\cal H}-\lambda {\bf S}^2$ in order to restore
${\bf S}^2$ symmetry approximately.
The above ansatz for states with good total spin are examples of the 
use of {\it constraints} in mean field approaches, 
which necessarily raise the energy above the mean field minimum. 
One could also project the symmetry-unrestricted HF orbitals as discussed 
in Ref.\ \onlinecite{YanP}. In the latter case, however,
the wave function is no longer a single Slater determinant but rather
a sum of few determinants of the corresponding symmetry operator. 
Therefore, the ground state energy with the restored symmetry
is no longer bound by the mean-field minimum and thus can be 
closer to the absolute minimum imposed by the variational 
principle. 
   
With the above results, we conclude that the lowest HF solution in region 
{\small II} requires a simultaneous breaking of the
spin and space symmetries. Taking into account the results of the preceding 
sections we can say that singlet-triplet mixing in region {\small II} is an 
artifact of the HF solution. However,
space symmetry breaking in this region is a true physical mechanism
indicating an intrinsic structural change in the exact wave function, 
as supported by the roto-vibrational model discussed above. 
We believe this peculiar combination of artifact and physics
is due to the smallness of the configuration space for a two-electron
system.
We must also point out that at high $R_{\it mp}$ and/or $W_{\it mp}$ 
(regions {\small IV} and {\small V}) the HF prediction  
fails to match the results of the analysis given in Sec.\ IV.

In contrast to the exact results of Fig.\ \ref{fig2}
where a region (although narrow) with $m=-2$ corresponding to a singlet state appears
at large $W_{\it mp}$, 
the HF approximation predicts only one $(S_z=0)\to (S_z=1)$ transition 
as $W_{\it mp}$ increases. This can be understood as an 
overestimation of the exchange energy in the HF model which tends to favor 
spin alignment whenever orbital overlapping occurs, as it does in region
{\small III} with circular orbitals, as well as in regions {\small IV} and 
{\small V} with two-lobed orbitals.

\begin{figure}[t]
\centerline{\psfig{figure=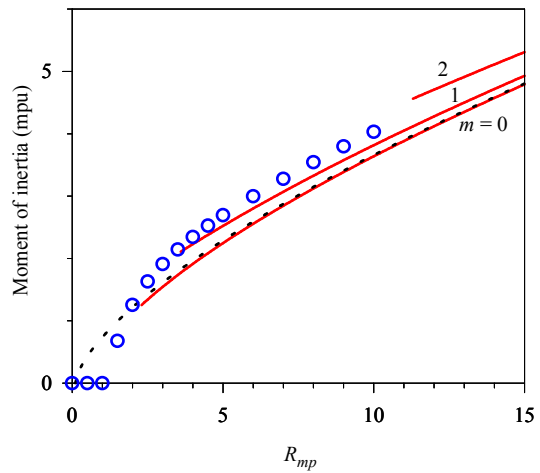,width=2.75in,clip=}}
\caption{(Color online) Moment of inertia computed in the RPA (circles) at 
$W_{\it mp}=0$ as a function of the adimensional parameter $R_{\it mp}$. Solid 
lines show the evolution of the corresponding values in the analytical model 
of Sec.\ III for different $m$-states. Each line starts at the 
crytallization onset for the corresponding angular momentum, according to
the criterion of Sec.\ IV.B. 
The dashed line represents the asymptotic 
value ${\cal J}=r_0^2/2$ taking $r_0\approx(2R_{\it mp})^{1/3}$, to which all 
solid lines converge at very high values of $R_{\it mp}$.}
\label{fig8}
\end{figure}

We consider next the results obtained by solving the RPA for excitations.
As discussed in Ref.\ \onlinecite{rpa} the RPA determines the 
moment of inertia associated with the collective rotation of a 
deformed HF structure (see Eq.\ (18) in Ref.\ \onlinecite{rpa}). 
Figure \ref{fig8} presents the evolution with $R_{\it mp}$, at $W_{\it mp}=0$, 
of the RPA moment of inertia ${\cal J}_{RPA}$ (circles). For comparison,
the values computed through 
the solution of Eq.\ (\ref{eq8}) of the roto-vibrational model,
${\cal J}=r_0^2/2$, are also shown (solid lines).
Each ${\cal J}$-line starts at the crystallization onset for the
corresponding angular momentum using the criterion of Sec.\ IV.B.
Note that ${\cal J}_{RPA}$ remains null until the 
HF solution breaks rotational symmetry at $R_{\it mp} \simeq 1$, from there on
it reasonably agrees with the exact values, somehow averaging
the exact results for different $m$'s. 
The molecule stretching, yielding larger $r_0$'s (${\cal J}$'s)
as $m$ increases, is obviously outside the RPA. 
All ${\cal J}$ values slowly converge to a 
common result with increasing $R_{\it mp}$, i.e., to
an exact rigid-rotor behavior. For comparison, the dashed line 
represents the asymptotic value corresponding to $r_0\approx(2R_{\it mp})^{1/3}$.
The nice qualitative agreement between ${\cal J}_{RPA}$ and 
${\cal J}$ is giving additional support to the above conclusion that
space symmetry breaking in region {\small II} of the HF phase diagram 
indicates a genuine physical effect and, thus, it also supports the overall 
picture of a rotating Wigner molecule.
 
Although the RPA restores circularity on the deformed HF mean field,\cite{rpa} 
associated with the $L_z$ operator, it is not able to restore the symmetry 
related to ${\bf S}^2$, since the latter one is a two-body operator 
which is beyond the RPA treatment of broken symmetries.
A side effect of the HF spin artifact in region {\small II} is that when spin-flip
bosonic pairs are included in the RPA, generalizing our previous calculation,\cite{rpa}
the correlation energy in this symmetry-broken phase is badly overestimated, 
i.e., it is between two and three times larger than the exact value.  
This does not occur, however, in the circularly symmetric regions.

Focusing now on the RPA vibron states, we must also distinguish
excitations associated with spin and space (charge) degrees of freedom.
While the RPA qualitatively describes all excitations in regions {\small I} and
{\small III}, it fails for spin excitations in phase {\small II}. Obviously, this is
due to the HF spin artifact in this region. It also fails for space
excitations of phases {\small IV} and {\small V}.
Finally, we end this section by pointing out that, in spite of the 
spin artifact of phase {\small II}, in pure triplet phases RPA 
reproduces the exact spin precession frequencies known 
from the theory of magnetic resonance.\cite{Sli}
That is, a pure spin-flip state (precessional mode) is expected at the 
Larmor energy 
$\hbar\omega_{\cal L}=g^*\mu_B B=\frac{1}{2} g^* m^* W_{\it mp}$ mpu. 
Indeed, within the RPA, i.e., in the quasi-boson approximation,\cite{BR,rpa} 
one finds 
\begin{equation}
[{\cal H}, O^+_{\cal L}] = \hbar\omega_{\cal L}  O^+_{\cal L}\; ,
\end{equation}
with the vibron operator for the Larmor mode
\begin{equation}
\label{eq15}
O^+_{\cal L}=\frac{S_x+iS_y}{\sqrt{2\langle S_z\rangle}} \; .
\end{equation}
In Eq.\ (\ref{eq15}) $\langle S_z\rangle$ is the HF expectation value of the 
$S_z$ operator.
In fact, the Larmor mode at $\hbar\omega_{\cal L}$ appears whenever
the ground state has $\langle S_z\rangle\ne 0$
and it is normally the lowest excitation of the system. 

\section{Summary}

We performed a systematic study of the evolution of ground and excited 
states of two-electron quantum dots subject to an external magnetic field.
The analysis has been done in terms of magneto-parabolic units
and the associated parameters $(R_{\it mp},W_{\it mp})$ that give, respectively, 
the ratios of Coulomb interaction strength and cyclotron frequency to effective
confinement. The ground state calculations are summarized in a phase diagram that can be 
equally applied to a variety of confinements, magnetic fields and material 
parameters.

We suggested an analytical model for the interpretation of the
exact results, including roto-vibrational coupling and centrifugal
distortion (molecule stretching).  
Within this roto-vibrational model we proposed a criterion to determine 
the onset of Wigner crystallization based on the decoupling
of rotational and vibrational motions. For a 3{\%}-decoupling threshold 
we found that Wigner crystallization appears, for zero-angular-momentum states,   
when $R_{\it mp}$ exceeds a value $\simeq 2$. 
States with larger $m$'s crystallize 
at higher $R_{mp}$ values.
In agreement with the homogeneous gas situation the potential energy 
of the crystallized states is much larger than the kinetic energy, the latter
one being solely due to the vibrational zero-point motion of the electrons.

The HF calculations predict that crystallization for $S_z=0$ occurs when 
$R_{\it mp}>1$, the new phase ({\small II}) being in an artificial mixture of 
singlet and triplet spin states. The space symmetry breaking in phase {\small II} is
a genuine physical effect but the spin mixture is an artifact due to the 
smallness of the configuration space for a two-electron system. Other HF 
symmetry-broken phases ({\small IV} and {\small V}) do not agree with the exact results.
The RPA moment of inertia qualitatively
agrees with the result from the roto-vibrational model, although 
the molecule centrifugal distortion is missed. 
On the other hand, the RPA produces reliable results for space (charge) excitations
in regions {\small I}, {\small II} and {\small III}, as well as for spin excitations when the 
HF solutions possess good ${\bf S}^2$ and $S_z$ quantum numbers (regions {\small I} 
and {\small III}).
We would expect a broader applicability of the many-body theories (HF+RPA) for
larger systems. Work along this line is in progress.
In conclusion, the combined use of exact and model calculations
allowed us to ascertain the existence of a rotating Wigner molecule
in a two-electron dot for relatively large electron densities or,
equivalently, small $R_{\it mp}$ parameters.

\begin{acknowledgments}
This work was supported by Grant No.\ BFM2002-03241 
from DGI (Spain). R. G. N. gratefully acknowledges support from the 
Ram\'on y Cajal programme (Spain).
\end{acknowledgments}



\begin{thebibliography}{99}
\bibitem{ana} 
              Strictly speaking, this expression is not fully analytical
since $r_0$ is determined by the algebraic Eq.\ (\ref{eq8}) that, in general, 
we solve numerically. Nevertheless, to simplify the discussion we shall 
continue to call this result and also, by extension, the complete 
roto-vibrational model, analytical.
\end{thebibliography}
\end{document}